\newcommand{\CLASSINPUTtoptextmargin}{2cm}
\newcommand{\CLASSINPUTbottomtextmargin}{2,9cm}

\documentclass[conference, a4paper]{IEEEtran}

\usepackage{cite}
\usepackage{bm}
\usepackage{amsmath,amssymb,amsfonts}
\usepackage{algorithmic}
\usepackage[utf8]{inputenc}
\usepackage{graphicx}
\usepackage{textcomp}
\usepackage{booktabs}
\usepackage{multirow}
\usepackage{longtable}
\usepackage{xcolor}
\usepackage{xspace}
\usepackage{multirow}
\usepackage{nopageno}
\usepackage{comment}

\usepackage{dblfloatfix}    

\newtheorem{definition}{Definition}
\usepackage[utf8]{inputenc}
\newcommand*{\TPC}{\texttt{TPCx-V}}

\usepackage{glossaries}

\newacronym[plural=CPUs,firstplural=Central Processing Units (CPUs)]{cpu}{CPU}{Central Processing Unit}
\newacronym[plural=HVMs,firstplural=Hardware Virtual Machines (HVMs)]{hvm}{HVM}{Hardware Virtual Machine}
\newacronym[plural=IDSes,firstplural=Intrusion Detection Systems (IDSes)]{ids}{IDS}{Intrusion Detection System}
\newacronym[plural=LKMs,firstplural=Loadable Kernel Modules (LKMs)]{lkm}{LKM}{Loadable Kernel Module}
\newacronym[plural=OSes,firstplural=Operating Systems (OSes)]{os}{OS}{Operating System}
\newacronym[plural=vCPUs,firstplural=Virtual CPUs (vCPUs)]{vcpu}{vCPU}{Virtual CPU}
\newacronym[plural=VMs,firstplural=Virtual Machines (VMs)]{vm}{VM}{Virtual Machine}

\newacronym{api}{API}{Application Programming Interface}
\newacronym{ba}{$BA$}{Bucket Algorithm}

\newacronym{cisuc}{CISUC}{Center for Informatics and Systems of the University of Coimbra}
\newacronym{ct}{CT}{Compromised Tenant}
\newacronym{db}{DB}{Dependability Benchmark}
\newacronym{dos}{DoS}{Denial-of-service}
\newacronym{dei}{DEI}{Department Informatics Engineering}
\newacronym{gdt}{GDT}{Global Descriptor Table}
\newacronym{iaas}{IaaS}{Infrastructure as a Service}
\newacronym{kvm}{KVM}{Kernel-based Virtual Machine}
\newacronym{mafalda}{MAFALDA}{Microkernel Assessment by Fault injection AnaLysis and Design Aid}
\newacronym{mcdm}{MCDM}{Multi-Criteria Decision Making}
\newacronym{oltp}{OLTP}{Online Transaction Processing}
\newacronym{pb}{PB}{Performance Benchmark}
\newacronym{pv}{PV}{Paravirtualization}
\newacronym{qa}{QAs}{Quality Attributes}
\newacronym{qm}{QM}{Quality Model}
\newacronym{rq}{RQ}{Research Question}
\newacronym{sb}{SB}{Security Benchmark}
\newacronym{spec}{SPEC}{System Performance Evaluation Cooperative}
\newacronym{sut}{SUT}{System Under Test}
\newacronym{tm}{TM}{Test Manager}

\newacronym{tpcxv}{\TPC}{TPC Express Benchmark V~\cite{TPCx-V_spec}}
\newacronym{tpc}{TPC}{Transaction Processing Performance Council}
\newacronym{vi}{VI}{Virtualization Infrastructures}
\newacronym{vmi}{VMI}{Virtual Machine Introspection}
\newacronym{vmm}{VMM}{Virtual Machine Monitor}
\newacronym{wg}{WG}{Working Group}
\newacronym{wsdl}{WSDL}{Web Services Description Language}

\glsresetall

\begin{document}

  \author{
    \IEEEauthorblockN{Charles F. Gon\c{c}alves\IEEEauthorrefmark{1}\IEEEauthorrefmark{2}, Daniel S. Menasché\IEEEauthorrefmark{4}, Alberto Avritzer\IEEEauthorrefmark{3}, Nuno Antunes\IEEEauthorrefmark{1}, Marco Vieira\IEEEauthorrefmark{1}}
    \IEEEauthorblockA{\IEEEauthorrefmark{1}University of Coimbra, CISUC, DEI, Portugal - 
     {\{charles,nmsa,mviera\}@dei.uc.pt}}
     \IEEEauthorblockA{\IEEEauthorrefmark{2}Information Governance Secretariat, CEFET-MG, Brazil -  
     charles@cefetmg.br
     }
    \IEEEauthorblockA{\IEEEauthorrefmark{3}eSulab Solutions, Princeton, New Jersey - beto@esulabsolutions.com}
    \IEEEauthorblockA{
\IEEEauthorrefmark{4}Federal University of Rio de Janeiro, Brazil  - 
sadoc@dcc.ufrj.br}
 }

\title{A Model-Based  Approach   to    Anomaly Detection \\ Trading  Detection Time and False Alarm Rate}

\thispagestyle{plain}
\pagestyle{plain}

\maketitle

\begin{abstract}
The complexity and ubiquity of modern computing systems is a fertile ground for anomalies, including security and privacy breaches.  
In this paper, we propose a new methodology that 
addresses the practical challenges to implement anomaly detection approaches. Specifically, it is challenging to define normal behavior comprehensively  and to acquire data on anomalies in diverse cloud environments.
To tackle those challenges, we focus on anomaly detection approaches based on system performance signatures. 
In particular, performance signatures have the potential of detecting zero-day attacks, as those approaches are based on detecting performance deviations and do not require detailed knowledge of attack history. 
The proposed methodology leverages an analytical performance model and experimentation, and  allows to control the  rate of false positives in a principled manner.   
The methodology is evaluated using the TPCx-V workload, which was profiled during a set of executions using resource exhaustion anomalies that emulate the effects of anomalies affecting system performance. 
The proposed  approach was able to successfully detect the anomalies, with a low number of false positives (precision 90\%--98\%).

\end{abstract}


\section{Introduction}
\label{sec:intro}








Complex computing systems, such as  cloud solutions~\cite{forbes_workload2020}, are ubiquitous.   Such ubiquity, in turn, is a potentially fertile ground for security and privacy breaches~\cite{intel, digitalocean, wallschlager2017automated, gulenko2016evaluating}. 
Efficiently detecting and mitigating such attacks is an important step to counter the threat that they pose to the existing IaaS systems and, more broadly, to the virtualization culture that supports a significant fraction of today's systems~\cite{hayes2008cloud}.

The design of intrusion detection systems (IDSs) for detecting anomalies, such as zero-day attacks and advanced persistent threats (APTs)~\cite{GrottkeAMAA16} in virtualized environments, poses several domain-specific challenges~\cite{Chandola2009,milenkoski2015evaluating}. 
In particular, $(i)$ it is challenging to comprehensively define normal behavior in a  diverse cloud environment, $(ii)$ malicious attackers may adapt their behavior to fit the domain definition of  ``normal behavior'', and $ (iii)$ data availability on anomalies at cloud environments, which would be key for training, is hard to obtain~\cite{erlacher2018fixids, erlacher2018testing}.  
To tackle those challenges, we focus on anomaly detection approaches based on system performance signatures. 
In particular, \textbf{performance signatures have the potential of detecting zero-day attacks}~\cite{Chandola2009,milenkoski2015evaluating}, as those approaches are based on detecting performance deviations and do not require detailed knowledge of attack history~\cite{FernandesJr2019}.

In this paper, we propose a methodology for anomaly  detection based on   performance deviations caused by anomalies in complex virtualized systems.
\textbf{The proposed tuning of the anomaly detection mechanism leverages an analytical performance model and experimentation, and   allows to control the  rate of false positives in a principled manner}~\cite{Chandola2009}.
After a careful analysis of every kind of transaction in the target system, the methodology profiles the system operation under normal conditions for its key transactions.
Then, during system operations performance monitoring,  performance deviations from the baseline are reported as anomalies.

To validate the proposed methodology, we ran an extensive experimental campaign using the \texttt{TPCx-V} workload~\cite{TPCx-V_spec}, which is representative of a large virtualized infrastructure that supports a business that relies on transactional systems.
Fault injection was used to emulate the effects of anomalies, e.g., due to  attacks,  impacting system performance.
Experience and practice show that injecting the effects of faults and attacks is an effective way to check systems dependability~\cite{arlat1990fault, arlat2003comparison}.

The experiments showed the applicability and effectiveness of the proposed anomaly detection methodology.
In our experiments, it was possible to detect most of the performance deviations, with a low number of false positives (precision of 90\% and 98\% for the worst and best configurations).
In addition,  given the model-based nature of the solution, it is amenable to what-if analaysis so as to trade between the rate of false positives and detection time. 



In summary, the paper's contributions are the following:

$(i)$ An \textbf{analytical model to support anomaly detection designs}, which allows conducting principled parameterization. The model can be used to cope with the tradeoff between time to detect an anomaly and the rate of false alarms (Section~\ref{sec:ad_algo}).

$(ii)$ An \textbf{experimental assessment  of the methodology in practice} using a representative system. We established the feasibility of detecting anomalies based on non-intrusive user-level performance metrics that are available in production environments (Sections~\ref{sec:expEnvl} and \ref{sec:results}).

$(iii)$ A \textbf{model-driven principled mechanism design} that allows revisiting the experimental results and conduct what-if analysis to assess different performance metrics of the considered anomaly detection algorithms as a function of the parameterization (Section~\ref{ssec:modelbasedinsights}).

%
%
%



The remainder of the paper is organized as follows.  
Section~\ref{sec:related} covers related work, followed by our contributions in Sections~\ref{sec:ad_algo}-\ref{sec:results} as indicated in the summary of contributions above. Finally, Section~\ref{conc} concludes the paper.

\color{black}

\section{Related Work} 
\label{sec:related}



In this section, we revise  related  literature indicating how  the current work relates to  prior art.

\subsection{ Anomaly detection for cybersecurity} 


An approach for anomaly detection consists in running sequential hypothesis tests~\cite{ho2011fast,jung2004fast}.
In~\cite{jung2004fast},  sequential hypothesis tests are used for  the detection of malicious port scanners. The authors have developed a link between the detection of malicious port scans and the theory of sequential hypothesis testing. They have also shown that port scanning can be modeled as a random walk.  The detection algorithm matches the random walk to one of two  stochastic processes, which correspond to malicious remote hosts or to authorized remote hosts.  The approach considered in this paper is similar in spirit to that considered in~\cite{ho2011fast,jung2004fast}, as \emph{our analytical results are derived from a  birth-death Markov chain}. Such Markov chain can be interpreted as a random walk  through buckets which fill as the system degrades, and empty as the system recovers (see Section~\ref{sec:analytmodel}).



A number of previous works have considered anomaly detection approaches  using performance signatures~\cite{Avritzer2005, Avritzer2007, Cherkasova2009, avritzer2010monitoring}. 
%
In~\cite{Avritzer2007} an approach for the mitigation of worm epidemics in tactical Mobile Ad-Hoc Networks (MANETs) using performance signatures (response time) and software rejuvenation was introduced.
The work in~\cite{Cherkasova2009} introduced a framework that detects anomalous application behavior using regression-based models and application performance signatures. 
Then,~\cite{avritzer2010monitoring} builds on top of previous work on performance signatures~\cite{Avritzer2005, Avritzer2007, Cherkasova2009} and proposes an anomaly detection approach based on performance signatures based on CPU, I/O, memory and network usage   for the detection of security intrusions. 

\subsection{Bucket algorithm and sequential decision making}

The performance of signature-based intrusion detection systems relies on  intrusion detection algorithms that account for workload variability  to avoid a high rate of false positive alerts.  An example of such workload-sensitive algorithms is the \gls{ba}  that was introduced in~\cite{Avritzer2005} and is presented in detail in Section~\ref{sec:ad_algo}.

In Section~\ref{sec:ad_algo}  
we revisit the \gls{ba} mechanism, and present an analytical model that is instrumental to parameterize the \gls{ba} from experimental data.
A statistical analysis  of the behavior of a   family of \glsplural{ba} has been described on~\cite{AvritzerBGTW06},  without accounting for the tradeoff between detection time  and false alarm rate. \emph{One of the goals of this paper is to fill that gap. }
In the previous cited research
~\cite{Avritzer2005, Avritzer2007},
simulations were used to support the  analysis of the \glsplural{ba} algorithms. In contrast, in this work we introduce an analytical  model of the
\glsplural{ba} algorithms that can be used
to  support  anomaly  detection designs, and an experimental assessment  of the  methodology in  practice using  a  representative  system.

\section{Anomaly Detection Mechanism and Model} 
\label{sec:ad_algo}

In this section we describe the anomaly detection mechanism considered in this paper  followed by the proposed  analytical model. 

\subsection{Anomaly Detection Mechanism} 

\begin{figure}[h]
    \includegraphics[width=1\linewidth]{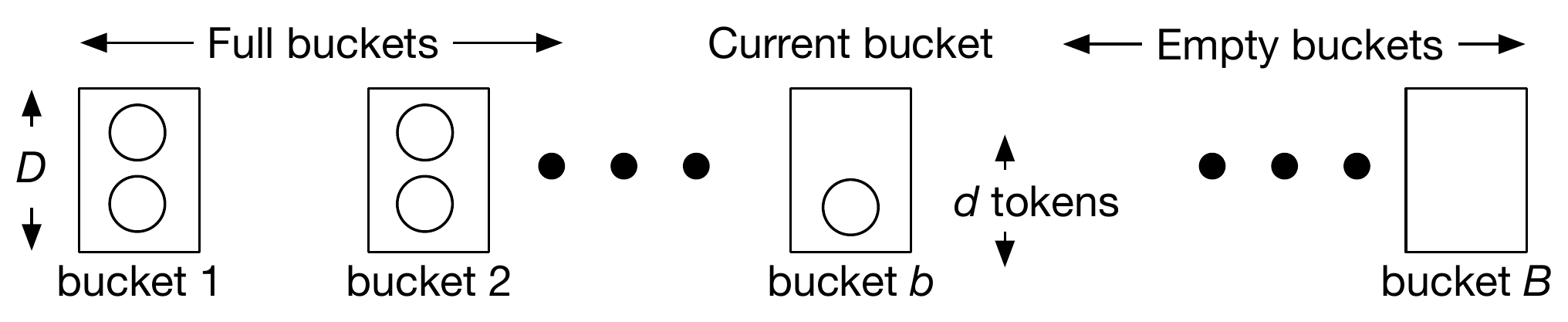} 
    \caption{System of buckets  representing the dynamics of the anomaly detection algorithm. 
    $D$ and $B$ must be properly parameterized for adequate operation.}
    \label{intrusion}
\end{figure}

The bucket algorithm for anomaly detection  based on performance degradation works by continuously measuring the throughput, $\overline{x}$, and maintaining 
$B$ buckets of depth $D$ each. 
Samples are added to and removed from buckets as a function of  the  history of  most recent throughput measurements, as shown in \figurename~\ref{intrusion}, wherein each ball corresponds to a throughput sample. 
The scalar value  $b$ is a pointer to the
current bucket, $b=1, \ldots, B$ and $d$ is the number of recent throughput samples stored in the current bucket, $d=0,1,\ldots,D$. 

Let $\mu$ be the baseline average throughput, and  $\sigma$ be the baseline standard deviation. Both $\mu$ and $\sigma$ can be  
derived from the execution of controlled  experiments without anomalies (i.e. \textbf{golden runs}). The pointer $b$ to the current bucket is used to determine the current target throughput, which is given by  $\overline{x} = \mu - (b-1)  \sigma$. Once the current bucket overflows (resp., underflows), the   target throughput  is shifted upward (resp., downward) by one standard deviation.  When all buckets overflow the algorithm detects a performance degradation and triggers an anomaly alarm. 
%
The performance degradation detection algorithm, that we will refer hereinafter as  Bucket Algorithm (\gls{ba}), is given as follows:

\emph{Initialization:} $\{b\leftarrow 1; d\leftarrow 0 \}$, with all buckets empty.
 
\emph{Main loop:} for each sample $\hat{x}$ of throughput, execute the steps below.
    
\begin{enumerate}
\item if $\hat{x} < (\mu - (b-1) \sigma)$ then $\{ d \leftarrow d + 1 \}$,  throughput smaller than reference value, add token to  current bucket;
\item else do $\{ d \leftarrow d - 1\}$,  throughput larger or equal than reference value, remove token from the current bucket;
\item if $(d > D)$ then do $\{ d \leftarrow 0; b \leftarrow b+1 \}$, current bucket overflow, go to next bucket;
\item if $\big((d < 0)$ and $(b >1)\big)$ then
do $\{ d \leftarrow D; b\leftarrow b-1\}$,  current bucket underflow,  go to previous bucket;
\item  if $\big((d < 0)$ and  $(b == 1)\big)$ then
do $\{d \leftarrow  0\}$ all buckets empty, system recovered from transient performance degradation;
\item  if $b > B$, all buckets overflow, trigger performance degradation alarm.
\end{enumerate}

The performance degradation detection algorithm can be tuned by varying the bucket depth, $D$, and the number of buckets, $B$.  The larger the product $D\times B$  the smaller the rate of false alarms but the longer it takes  for the algorithm to detect the performance degradation.


 
%
%

\subsection{Hypothesis Testing}

The system administrator continuously considers two alternative hypothesis:  $(i)$ null hypothesis $H_0$ corresponding to a situation where there is no anomaly  taking place
and $(ii)$ alternative hypothesis $H_1$ meaning that there is an anomaly, e.g., the system is under attack. Then, the key quantities of interest can be defined as a function of $H_0$ and $H_1$.
   To simplify presentation, in what follows time is measured in number of collected samples.
\begin{definition} \label{def:1}
The  mean time until a false alarm under $H_0$ is denoted by $A_B(D)$. 
\end{definition}

As discussed in the following section,  $A_B(D)$ is given by the mean time to  reach the absorbing state of a Markov chain characterizing the bucket algorithm. When $B=2$, we provide closed-form  expressions for $A_B(D)$.


\begin{definition}  \label{def:2}
A lower bound on the  number of samples until a true positive  under $H_1$  is denoted by $L$.  Assuming all buckets are initially empty, we let $
    L = BD$.
\end{definition}

\begin{definition}  \label{def:3}
The probability of false alarm under $H_0$ is the probability that an alarm is triggered outside an anomaly,
$  f_B(D) = \mathbb{P}(R < T), $ 
where $R$ is a random variable with mean $A_B(D)$ characterizing the time until an alarm is triggered,  and  $T$ is  a random variable with mean $1/\alpha$ characterizing the time until an anomaly occurs.
\end{definition}

In this paper, except otherwise noted we assume that $f_B(D)$ depends on $R$ and $T$ only through their means.


\begin{definition} \label{def:cost}
The expected cost of a given system parameterization is a weighted sum of the probability of false alarms, computed under $H_0$,  and a lower bound on the number of samples to detect an anomaly, computed under $H_1$,
 \begin{equation}
     C(\bm{p}, w, D, B, \alpha)  =  B D +  w f_B(D).  \label{eq:cost1}
 \end{equation}
\end{definition}

Table~\ref{tab:tabnotation} summarizes the notation introduced in this section. 
Additional details about how  to estimate $A_B(D)$ and $f_B(D)$ are provided in Sections~\ref{sec:analytmodel} and~\ref{sec:modelfa}, respectively. Then, the cost function~\eqref{eq:cost1} (Definition~\ref{def:cost}) will be instrumental to parameterize the bucket algorithm  in Section~\ref{sec:optimization}.

\begin{table}[h]
    \centering
        \caption{Table of notation}
         \vspace{-0.1in}
    \begin{tabular}{l|l}
        \hline
        variable & description  \\
        \hline
        $B$ & number of buckets \\
        $b$ & current bucket, $b=1, \ldots, B$ \\
        $D$ & maximum bucket depth \\ 
        $d$ & current depth of bucket $b$, $d=0, \ldots, D$ \\
        $A_B(D)$ & mean time to false alarm, under $H_0$ (no anomaly), i.e.: mean\\
        &  number of collected samples to reach absorbing state \\
        $f_B(D)$ & probability of false alarm \\
        $F$ & target probability of false alarm \\
        $\alpha$ & anomaly rate \\
        $p_i$ & probability that  sample adds  ball to bucket,  when  $b=i$   \\
         \hline
    \end{tabular}
    \label{tab:tabnotation}
\end{table}

\subsection{Analytical Model}
\label{sec:analytmodel}
Simple algorithms to detect anomalies, such as the bucket algorithm, can be tuned using first principles.
%
%
%
%
The larger the depth of the bucket, the lower the false alarm probability, but the longer it takes for a true positive to be identified.
To simplify the analysis, we \textbf{work under the assumption} that anomalies will change the throughput distribution, and will always be detected.
However, the number of samples to detect the anomaly may vary depending on the depth of the bucket.
Our \textbf{second key simplifying assumption is} that the number of samples to detect the anomaly is much smaller than the number of samples collected before getting a false alarm.  \emph{The two assumptions above are mild, and should typically hold in real settings} as the time until a false alarm in practical systems should be much longer on average than the time until a true positive~\cite{GrottkeAMAA16,Avritzer2007}.
Then, we aim at answering the following question: what is the smallest bucket depth to produce a false alarm probability upper  bounded by a given threshold?

Next, we introduce a 
discrete time birth-death Markov chain (DTMC) to characterize the behavior of the \gls{ba}. State $(b,d)$ of the Markov chain corresponds to the setup wherein there are $d$ balls in bucket $b$, and $D$ balls in buckets $b-1, \ldots, 1$.

\begin{figure}[!h]
    \centering
    \includegraphics[width=1\columnwidth]{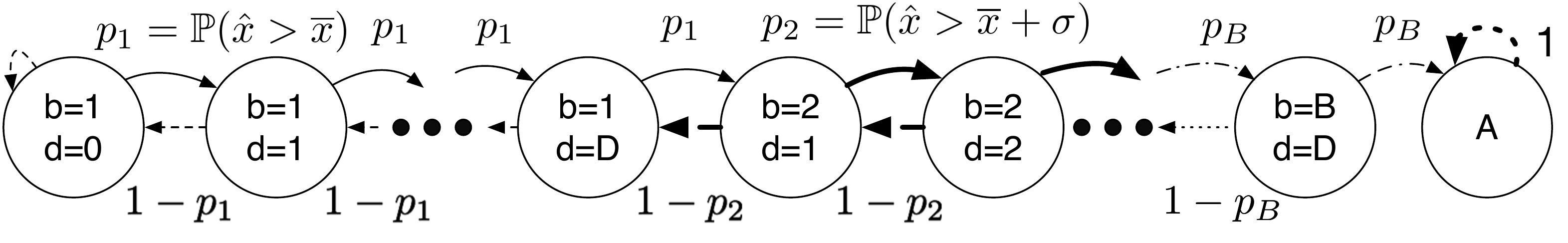}
    \caption{Discrete time Markov chain characterizing the behavior of the \gls{ba}.  Each transition corresponds to the collection of a new sample.}
    \label{fig:dtmc}
\end{figure}

Each  transition of the DTMC corresponds to the collection of a new sample. 
Such sample   causes the system to transition from state $(b,d)$ to one of its two neighboring  states. 
Let  $p_i$  be the probability that the number of balls at bucket $i$ increases after a new sample is collected.  
Then, $p_i=\mathbb{P}(\hat{x}  > \overline{x} + (i-1)\sigma),$ for  $ 1 \leq i \leq B$. 
The entries of the transition probability matrix are readily obtained from \figurename~\ref{fig:dtmc}.

Once the terminal absorbing state  is reached  an alarm is triggered (state $A$ in Figure~\ref{fig:dtmc}). 
The number of samples collected until absorption  accounts for a tradeoff between the mean time until $(a)$ a  false alarm, in the absence of anomalies, and $(b)$ detection, in the presence of an anomaly.  Larger values of bucket depth  $D$ favor the reduction of the former but  increase the latter.

Let $\tilde{A}_B(D;p_1,p_2)$ be  the  time until absorption, measured in number of collected samples, accounting for $B$ buckets of depth $D$ each. We denote its mean by $A_B$,    $\mathbb{E}(\tilde{A}_B)=A_B$.
Under the hypothesis of no anomaly,  $\tilde{A}_B$  is  the  time to a false alarm.  
We derived in~\cite{techreport} a closed-form  expression for $A_B$, which  is instrumental to handle  tradeoffs in the choice of the bucket depth $D$ as illustrated in the upcoming sections.
In particular, for $B=2$, the resulting expression is given by
%
\begin{align}
A_2(D;p_1,p_2)  
&= A^{(1)}_2(D; p_1,p_2)  + A^{(2)}_2(D; p_1,p_2) \label{eq:a2p1p2d}
\end{align}
where $A^{(i)}_B$ is the mean time to fill the $i$-th out of $B$ buckets,
\begin{align}
A^{(1)}_2 &=   { \Delta_1} \left(\delta_1- D\right) \\
A^{(2)}_2 
&=  \Delta_2\left( \delta_2
-D\right) + \Delta_1\left(\frac{1-\rho_1^{D+1}}{\rho_1^{D+1}}\right)  \delta_2
\end{align}
and
\begin{equation}
    \rho_i = \frac{1}{p_i} -1, \qquad \Delta_i=\frac{1+\rho_i}{1-\rho_i}, \qquad \delta_i = \frac{1-\rho_i^{-D}}{\rho_i-1}.   \label{eq:final2}
\end{equation}
Our experimental results indicate that $B=2$  suffices in the considered scenarios (see Section~\ref{sec:expEnvl}). For this reason, in the remainder of this paper all numerical results derived from the proposed analytical model are reported letting  $B=2$, making use of  equations~\eqref{eq:a2p1p2d}-\eqref{eq:final2}.   In what follows, we illustrate how to leverage the proposed model to estimate the probability of false alarms.

\subsection{Modeling the Probability of False Alarms}  \label{sec:modelfa}
Next, we  leverage the proposed model to estimate the probability of false alarms.    To that aim, we assume that anomalies, e.g., due to attacks, arrive according to a Poisson process with rate $\alpha$. 
Recall that $f_B(D)$ denotes the probability of a false alarm  (Definition~\ref{def:3}). 
In what follows, we derive expressions for $f_B(D)$ under different assumptions on the distribution of $\tilde{A}_B(D)$. 

Assuming that $\tilde{A}_B(D)$ can be roughly approximated by a constant, and that the  time between anomalies is exponentially distributed with mean $1/\alpha$,
\begin{equation}
    f_B(D) = e^{- A_B(D) \alpha}. \label{eq:fexp}
\end{equation}
Alternatively, if we approximate $\tilde{A}_B(D)$ by an exponential distribution,
\begin{align} \label{eq:fexpexp} 
    f_B(D) &= \frac{1/A_B(D)}{1/A_B(D) + \alpha} =  \frac{1}{1+{A_B(D)} \alpha }.
\end{align}
In the expressions above, we made  the dependence of $f_B$ and $A_B$ on the  bucket depth $D$ explicit as one of our goals is to study the relationship between $D$, $f_B$ and $A_B$.    
The closed-form equations~\eqref{eq:fexp} and~\eqref{eq:fexpexp} are instrumental to get insights about the interplay between the different model parameters.  In particular, as $D$ increases $A_B$ increases and $f_B$ decreases (Definition~\ref{def:1}), 
but the time to detect an anomaly increases  (Definition~\ref{def:2}).  
As indicated in the sequel, the equations above allow us to  find the minimum $D$ such that $f_B(D)$ is below a given threshold.  Then, in Section~\ref{sec:expEnvl} we experimentally validate that the values of $D$ obtained through the proposed model  produce the desired probability of false alarms in realistic settings.

\subsection{Parameterization of the Anomaly Detection Mechanism:  a Model-Driven Optimization Approach} \label{sec:optimization}

Next,   we show how to use the proposed model and the obtained expressions of  probability of false alarm for the purposes of  running  statistical hypothesis tests to determine whether there is an ongoing anomaly in the system.

Given a target false alarm probability, denoted by $F$, the system administrator goal is to  determine the optimal number of buckets and bucket  depth so as to 
minimize the lower bound on number of samples to detect an anomaly, $L$, while still meeting the target false alarm probability. 
\begin{align}
&{\mbox{\sc  Problem with Hard Constraints:}}\quad  \nonumber \\
&    \min \quad L=BD \\ \label{eq:hard}
 &   \textrm{subject to} \quad f_B(D) \leq F  
\end{align}
In what follows, we assume that $B$ is fixed and given.  Then, as $f_B(D)$ is strictly decreasing with respect to $D$, the constraint above will be always active and the problem translates into finding the minimum value of $D$ satisfying the constraint.  The problem above is similar in spirit to a Neyman-Pearson hypothesis test, for which similar considerations apply, i.e., the optimal parameterization of the test is the one that satisfies a constraint on the false alarm probability.  

Alternatively, the problem above can be formulated through the corresponding Lagrangian, 
\begin{align}
&{\mbox{\sc  Problem with Soft Constraints:}}\quad \nonumber  \\
& \min     \mathcal{L}(D) = B D + w (f_B(D)- F)  \label{eq:lag} 
\end{align}
where $w$ is the Lagrange multiplier.  The Lagrangian naturally leads to an alternative formulation of the problem, wherein  the hard constraint in~\eqref{eq:hard} is replaced by a soft constraint corresponding to the penalty term $f_B(D)- F $ present in the cost Lagrangian. The Lagrangian is a cost function, motivating Definition~\ref{def:cost}. Note that as $wF$ is a constant, minimizing~\eqref{eq:lag} is equivalent to minimizing~\eqref{eq:cost1}.

\section{Experimental  Setup and Fault Model } 
\label{sec:expEnvl}

To illustrate and validate the methodology presented in Section~\ref{sec:ad_algo}, we ran an experimental campaign using the \gls{tpcxv}, as described in Sections~\ref{ssec:sut} and \ref{ssec:expsetup}. 
%
%
%
%
A fault injection approach was used to emulate the effects of performance affecting security intrusions, as described in Section~\ref{ssec:faults}.  Then, the  model-based calibration of the anomaly detector is reported in Section~\ref{ssec:modelbasedinsights}.  

\subsection{System Under Test} 
\label{ssec:sut}


The \gls{tpcxv} is a publicly  available, end-to-end benchmark for data-centric workload on virtual servers. The benchmark kit provides the specification,  implementation, and tools to audit and run the benchmark. Details can be found in~\cite{tpcv2013, tpcv2015}.
 \gls{tpcxv} models many features commonly present in cloud computing environments such as multiple \glspl{vm} running at different load demand levels, and significant fluctuations in the load level of each \gls{vm}~\cite{tpcv2013}.  
We use the workload and software provided by the \gls{tpcxv}\cite{tpcv2015} to emulate a context closely related to a real-world scenario of brokerage firms that must manage customer accounts, execute customer trade orders, and be responsible for the interactions of customers with financial markets.

The goal of \gls{tpcxv} is to measure how a virtualized server runs database workloads, using them to measure the performance of virtualized platforms, specifically the hypervisor, the server hardware, storage, and networking. The minimal deployment of the \gls{tpcxv} 
comprises four groups of three \glspl{vm}, representing four different subsystems.   Table~\ref{tab:vms_spec} summarizes the considered experimental setup.   In Table~\ref{tab:vms_spec}, tpc-g\textit{n} refers to a \gls{vm} of group \textit{n}. Each group  was defined according to  the benchmark recommendations~\cite{TPCx-V_spec}.  

%
%
The \gls{tpcxv} workload is made up of \textbf{\texttt{12} types of transactions} 
that are submitted for processing at multiple databases (market, customer, and broker) following a specified mix of transactions per load  phase.  A typical run consists of  {\bf{\texttt{10} distinct load phases} of \texttt{12} minutes each}. 
 Transactions  simulate the stock trade process. When a trade finishes, a transaction named  \textit{Trade-Result} is issued.  
The primary performance metric for the benchmark is the business throughput (\texttt{tpsV}). It represents the number of completed \textit{Trade-Result} transactions per second.


\subsection{Experimental Setup} 
\label{ssec:expsetup}
Our \textit{experimental setup} is a deployment of the \gls{tpcxv} over two physical servers. The first server is a Dell PowerEdge R710 with 24 Cores, 96 GB RAM, and 12 TB disk, and is managed by a Xen hypervisor (4.4.1). It has a privileged domain (dom0) with a dedicated VM, and 15 additional VMs. One  set  of 12 VMs is dedicated to \gls{tpcxv}, with four groups of three VMs each  (\textit{tpc-xxx}). Another set of 3 VMs corresponds to an additional group running the compromised system. 
The second server (2 Cores, 8GB RAM and 1 TB disk) runs the same software and a single VM used for the driver component as prescribed on the \gls{tpcxv} specification. The details of each \gls{vm}  are described in Table~\ref{tab:vms_spec}.


\begin{table}[t]
\centering
\caption{ \scriptsize Experimental setup } 
\label{tab:vms_spec}
\vspace{-0.05in}
\setlength\tabcolsep{2pt} 
\begin{tabular}{lrr||lrr||lrr}
\toprule

\textbf{\gls{vm} } & \textbf{MB}  & \textbf{vCPU}   & \textbf{\gls{vm}} & \textbf{MB}  & \textbf{vCPU}  & \textbf{\gls{vm}} & \textbf{MB}  & \textbf{vCPU}     \\ \midrule
tpc-g1a         &   1024    & 1  & 
tpc-g1b2        &   4096    & 4  & 
tpc-g1b1        &   8192    & 2  \\
tpc-g2a         &   1024    & 1  & 
tpc-g2b1        &   12288   & 2  &
tpc-g2b2        &   6144    & 6  \\ 
tpc-g3a         &   1024    & 1  & 
tpc-g3b1        &   16384   & 2  & 
tpc-g3b2        &   8192    & 8  \\ 
tpc-g4a         &   1024    & 2  &
tpc-g4b1        &   20400   & 2  &
tpc-g4b2        &   10240   & 8  \\
\midrule
tenant A 		&   1532    & 2  & 
tenant B 		& 1532      & 2  & 
tenant C        &   1532    & 2 \\ 
\midrule
dom0            &   1962    & 4  & 
tpc-driver   &   1882    & 2 & & &\\ 
\bottomrule
\bottomrule
\end{tabular}
\end{table}



\subsection{Tools Set and Fault Model} 
\label{ssec:toolsup}

Each \textbf{single experiment lasts roughly 4 hours} comprising a  minimum of 2h as demanded by  the benchmark specification and additional 2h to track the  restoration of  the full environment to its initial state after performance degradation. Initial state restoration is achieved by rebooting the servers and recovering the system and databases, including the restoration of  all virtual disks.


\label{ssec:faults}

Next, we describe the considered  fault model.  Our goal is to account   for   resource exhaustion faults~\cite{Gruschka2010}.   
To that aim, we make use of  the Stress-NG module~\cite{stressng}, which was   designed to ``exercise various physical subsystems of a computer as well as the various operating system kernel interfaces''. 
In particular, we have defined three reference configurations to emulate  resource exhaustion anomalies, e.g., due to attacks.  

 \textit{(\texttt{H})}:  A high intensity workload consists of I/O intensive tasks executed by eight parallel processes   for 300 seconds.
 
 \textit{(\texttt{L})}:  A low intensity workload  comprises the execution of ten intervals, where each interval consists of 15 seconds of I/O intensive tasks executed by two parallel processes followed by  15   seconds in idle mode. 
  
 \textit{(\texttt{Ls})}: A shorter low intensity workload is similar to a low intensity workload  configuration except that it comprises the execution of   three intervals.




Recall from Section~\ref{ssec:sut} that  \gls{tpcxv} comprises 10 load phases with diverse load demands.  
We emulated resource exhaustion   anomalies on phases 4 and 6,  which correspond to  phases wherein the usage of physical resources and the reference \texttt{tpsV} reach their maximum values, respectively.  
%
%
Combining the observations above, we have a total of six fault models, which we will refer to using the corresponding phase number  followed by  the reference configuration: \texttt{4H}, \texttt{4L}, \texttt{4Ls}, \texttt{6H}, \texttt{6L}, and \texttt{6Ls}.
An execution of the system without fault emulation is referred to as a \emph{golden run}. The model based calibration of the anomaly detector introduced in the upcoming section is solely based on \emph{golden runs}, whereas the results presented in Section~\ref{sec:results} leverage the calibrated anomaly detector together with  the fault model introduced above.

\section{Model Based Calibration of Anomaly Detector and Counterfactual Analysis} 
\label{ssec:modelbasedinsights}

This section provides insights on the experimental results using the proposed model.  Our goals are to $(a)$ illustrate the applicability of the model in the real experimental setting described in the previous section; and $(b)$ indicate how the model can be used to trade between contending aspects such as false positive rate and time to detect anomalies.  

To exemplify the general process, we focus on one of the twelve transactions  referred to in Section~\ref{ssec:sut}, namely the  TRADE\_LOOKUP transaction, for which we  identify  that  $p_1=0.46$ and  $p_2=0.71$ considering the emulation of  the  system without anomalies (i.e., considering  golden runs). 
We assess the expected number of samples until a false alarm, obtained from~\eqref{eq:a2p1p2d}, letting  $B=2$ and $D$ vary between 1 and 30.  For $D=15$,  for instance,  the number of samples until a false alarm as estimated by the model already surpasses  $10^7$.   

\begin{figure}[t]
    \centering
    \includegraphics[width=\linewidth]{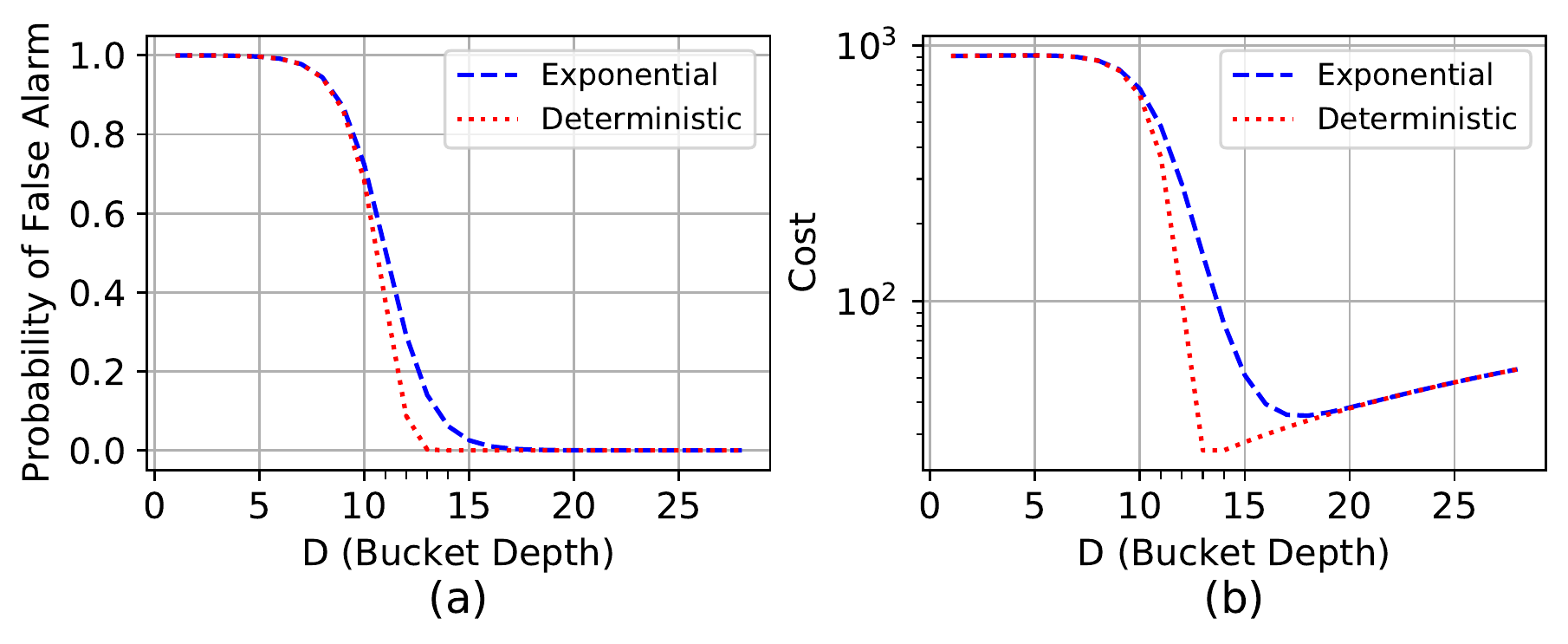} 
\caption{Model based calibration of anomaly detector: (a) probability of false alarm and (b) cost given by Equation~\eqref{eq:cost1} as a function of bucket depth.  }
\label{fig:modelpar2}
\end{figure}

\figurename~\ref{fig:modelpar2}(a) shows the probability of false alarm as a function of the bucket depth.   
\figurename~\ref{fig:modelpar2}(a) 
accounts for a fault model  wherein the mean time between  anomalies is $1/\alpha=5 \times 10^5$ samples. The dashed (resp., dotted) line corresponds to the exponential (resp., deterministic) approximation  for the time between anomalies, corresponding to~\eqref{eq:fexpexp} (resp.,~\eqref{eq:fexp}). 
As the bucket depth increases, the probability of false alarm decreases. 
For $D  \ge 12$, the probability of false alarm is  close to~zero.

As discussed in Section~\ref{sec:ad_algo}, there is a tradeoff between the probability of false alarm and the time to detect anomalies once they occur.  To cope with such a tradeoff, we consider both approaches  introduced in Section~\ref{sec:optimization}, namely the hard and soft constraint problems.  
Under the hard constraint problem,  a target probability of false alarm is determined, and the minimum value of $D$ that satisfies such target is sought.  For instance, if we set $F=0.03$ in~\eqref{eq:hard} then the minimum values of $D$ satisfying the constraint are $D=15$ and $D=13$ under the exponential and deterministic fault models, respectively.

Next, we assess  the cost  $C(\bm{p}, w, D, B, \alpha)$ introduced in Definition~\ref{def:cost}. Figure~\ref{fig:modelpar2}(b) shows how  the cost  varies as  a function of $D$,  letting $B=2$, $p_1=0.46$, $p_2=0.71$ and $\alpha=2 \times  10^{-6}$. 
To generate the plots, we let $w=909$, which corresponds to the Lagrange multiplier of the constrained problem under the deterministic model (see also~\eqref{eq:lag}).  In that setting,  the optimal bucket depth equals $D=13$ (see dotted line in Figure~\ref{fig:modelpar2}(b)), 
which is in agreement with the result  presented in the previous paragraph.    Note that for the exponential model the minimum cost is attained at $D=18$ (dashed line in Figure~\ref{fig:modelpar2}(b)), which is slightly larger than $D=15$ obtained in the previous paragraph. This is because the Lagrange multiplier corresponding to the exponential model is $w=57$.  Using such a smaller weight  favors a reduction in the optimal bucket depth to $D=15$, again in agreement with the results discussed in the previous paragraph.  


\emph{Take away message and engineering implications: }
the analysis presented in this section is instrumental to perform what-if counterfactual analysis and execute utility-driven model parameterization.  If the system administrator implements global countermeasures against attacks, for instance, it is expected that the rate of anomalies will decrease. In that case, the bucket depth can be adjusted accordingly using the approach introduced above.

The results presented in this section are also instrumental to reverse engineer the utility function subsumed by existing systems.  As indicated in the following section, letting $D$ vary between $12$ and  $15$ performed well  in the considered real scenarios.   The analysis presented above shows that    a system operating with $B=2$ and  $D=15$ is optimal, for instance, in case $5\times 10^{5}$ samples are collected inbetween anomalies and $w=57$, leading to a false positive probability of roughly 0.03.   Knowing that this is the case, one can tune the utility function, e.g., to account for anomalies that occur at different rates, and verify when/if the parameters of the bucket algorithm should be adjusted.

\color{black}

\section{Experimental Assessment of the Calibrated Anomaly Detector in Face of Faults}
\label{sec:results}

In what follows, we provide additional experimental evidence of the effects of the bucket depth $D$ on different system metrics. Motivated by the model-based analysis presented in Section~\ref{ssec:modelbasedinsights}, we focus most of our attention  on values of maximum bucket depth $D$ varying  between 12 and 15.
Our goals are to $(a)$   analyze the residual effects that can arise after the anomalies; and $(b)$ assess  the effectiveness of the proposed anomaly detection approach through two case studies and accounting for standard  performance metrics as detailed next. 

Our results are discussed using three metrics widely adopted in classification assessment~\cite{zaki2014dataminingbook}, namely  precision, recall and F-measure, which are defined as a function of true positives (TP), false positives (FP) and false negatives (FN) as follows,
\begin{align}
\textrm{Pr} = \frac{TP}{TP + FP},  \ \ &
\textrm{Re} = \frac{TP}{TP + FN},    &
\textrm{F1} = \frac{2 \times \textrm{Pr} \times \textrm{Re}}{\textrm{Pr} + \textrm{Re}}.   \nonumber
\end{align}
Precision (Pr) measures the impact of FP in the method's positive prediction.
Recall (Re)  reflects the sensitiveness of the algorithm, capturing the fraction of correct  predictions. 
 F-measure (F1) is the harmonic mean of precision and recall.
 
 
%


\begin{figure}[t]
\centering
\includegraphics[width=\linewidth]{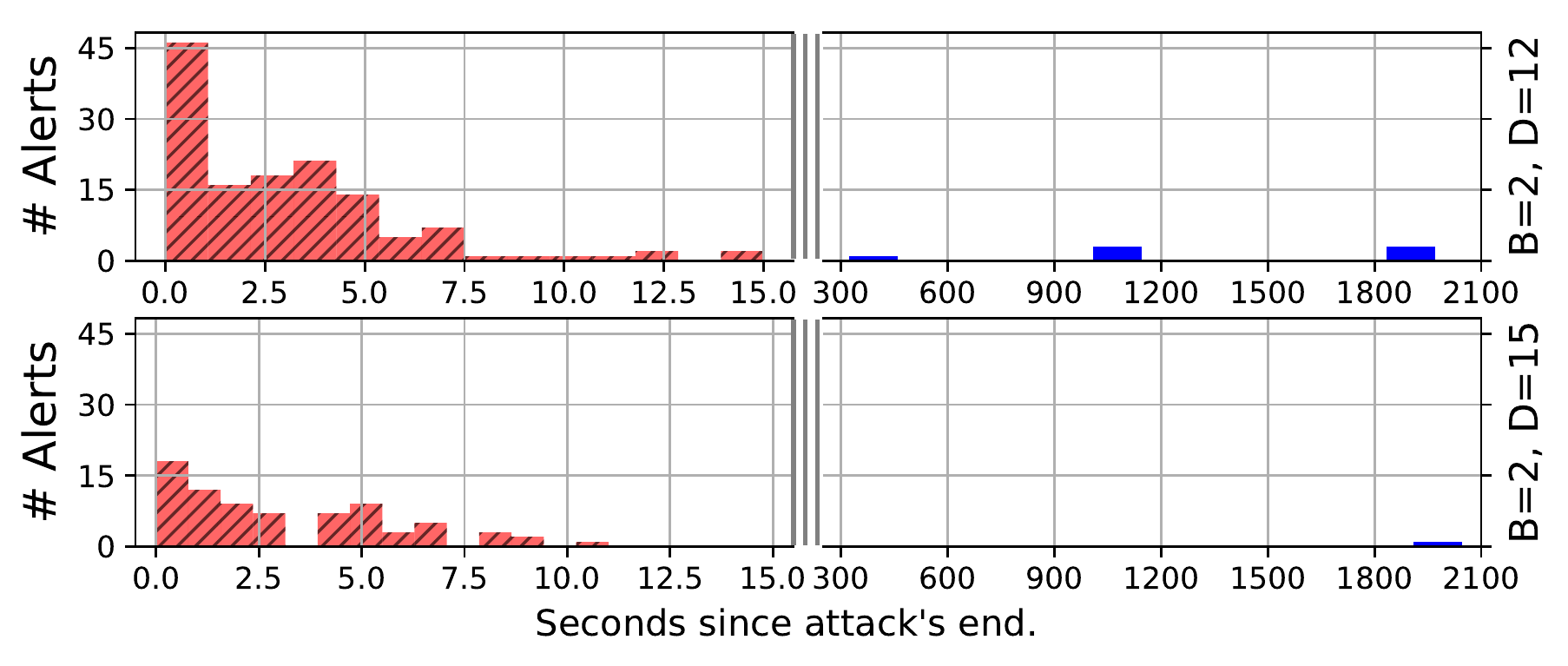}
\caption{Post-attack alerts distribution for bucket configuration with B=2 and D=[12,15]. Note that x scale is cropped to improve readability.}
\label{fig:re}
\end{figure}

\subsection{Residual Effects and Survivability Analysis}
\label{sec:re}

In our experiments we observed that the number of  bucket overflows after an attack ends was significantly greater  than in other non-attack periods.
This phenomenon is typically studied in the realm of survivability analysis, which focuses on system behavior from failure up to full recovery, including transient performance degradation~\cite{heegaard2009network, trivedi2006survivability}.
In this section,  we refer to the event of an  overflow of the $B$-th bucket as an \emph{alert}, and  explicitly distinguish alerts from alarms.  Intuitively, after an alarm is triggered during an attack, the set of alerts  caused by  transient effects  should  not trigger additional alarms.

%
%
\figurename~\ref{fig:re} shows  the number of alerts as  a function of time.   Following the survivability  perspective,  time is measured in seconds after an attack ends.  Note that a  significant fraction of alerts  occurs a few seconds after the attack, which suggests that those alerts are due to residual effects of attacks.   

Next, we propose a simple heuristic  to determine when a set of alerts  should be aggregated into a single alarm.  
To that aim, let $\delta$ be  the meantime until the first alarm is triggered during an attack phase.  
Table~\ref{tab:re_thrsd} reports how $\delta$ varies as a function of $D$ for two of the twelve transactions considered in our workload.  
The values of $\delta$ are relatively stable across  transactions and bucket depths. Note that $\delta$ increases as $D$ grows, up to $D=12$. Correspondingly, as $D$ grows the overall number of alerts decreases  (see Definition~\ref{def:2}). Together, Figure~\ref{fig:re} and Table~\ref{tab:re_thrsd} suggest the following heuristic:  after an alert at time $t_0$, any additional alerts during the interval  $[t_0, t_0 +c  \delta]$ are due to residual effects (e.g., emptying of queues and recovery of error states) and are discarded. In our experiments we let $c=3$.   It is also worth noting the slight decrease in $\delta$ when $D$ varies from 12 to 15.  We are currently investigating such a decrease, noting that it may not be statistically significant  
as the number of alerts decreases from $78$ and $88$ to $65$ and $71$, respectively, when $D$ varies from 12 to 15 for the two transactions in  Table~\ref{tab:re_thrsd}.
Together, Figure~\ref{fig:re} and Table~\ref{tab:re_thrsd} suggest the following heuristic:  any alert that occurs at time  $t < \delta$ is due to residual effects (e.g., emptying of queues and recovery of error states).   


\begin{table}[b]
    \centering
    \vspace{-0.2in}
    \caption{\scriptsize Mean time to first alarm ($\delta$) during the anomaly injection (in seconds)}
    \vspace{-0.1in}
    \label{tab:re_thrsd}
    \begin{tabular}{ccccc}
    \toprule
    \textit{\textbf{Transaction}} & \textbf{D=6} & \textbf{D=9} & \textbf{D=12} & \textbf{D=15} \\\midrule
    TRADE\_LOOKUP & 31.03 & 50.06 & 69.18 & 61.63 \\
    MARKET\_WATCH & 36.08 & 54.11 & 60.38 & 59.51 \\\bottomrule
    \end{tabular}
\end{table}

\begin{figure}[h]
    \centering
    \includegraphics[width=\linewidth]{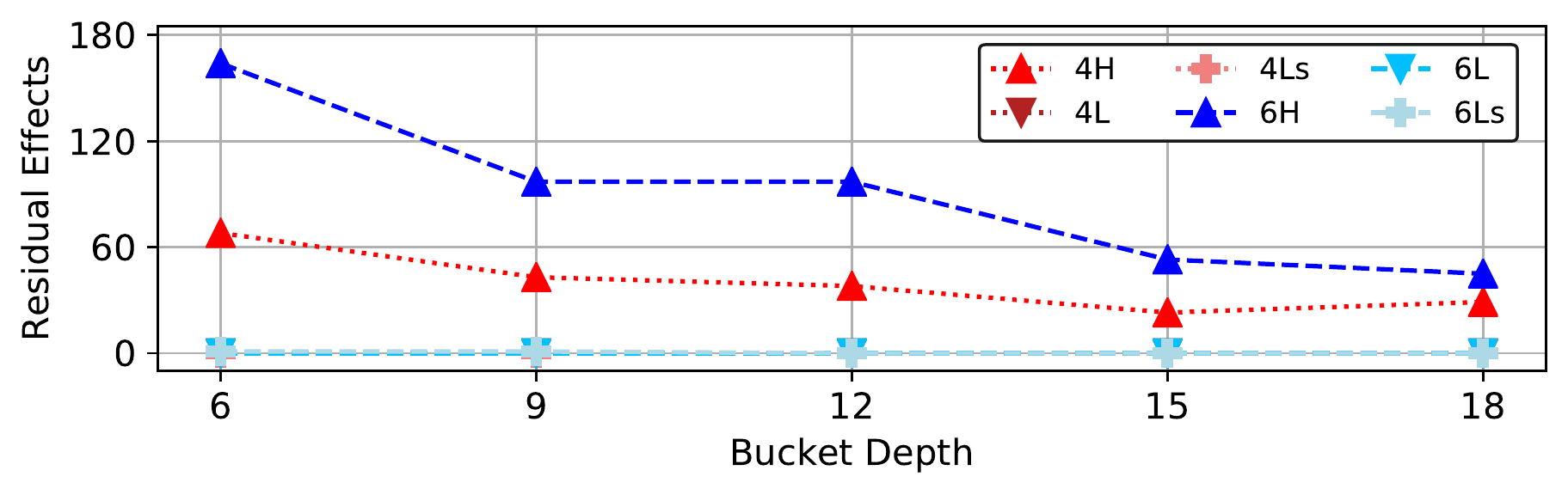}
    \vspace{-0.3in}
    \caption{Distribution of the residual effects by failure mode and bucket depth.}
    \label{fig:re_breakdwon}
    \vspace{-0.05in}
\end{figure}

\figurename~\ref{fig:re_breakdwon} shows the number of residual alerts as a function of $D$, for the six  fault models introduced in Section~\ref{ssec:toolsup}.  
The proposed heuristic yields   
residual alerts  under   high intensity faults (\texttt{6H} and \texttt{4H}, corresponding to blue dashed lines and red dotted lines).
In addition, the number of residual alerts decreases as 
 $D$ grows, as the 
the larger the value of  $D$  the higher is the tolerance for transient faults. 

Under the scenarios considered in this paper, the proposed heuristic \emph{accurately classified all  alerts due to residual effects  as spurious, and did not misclassify any non-residual alert and  the heuristic is subsumed under all the reported results  in the sequel. }

\subsection{ Experimental Assessment of  Parametrization Impact on Performance}

Next, our goals are to $(a)$  assess the performance  of the \gls{ba} over the six considered fault models and $(b)$ indicate how its parametrization affects  performance.

We consider the same bucket depth $D$ across all operations. Anomalies are detected per transaction and per VM group.  Noting that  9 out of the 12 transactions referred to in Section~\ref{ssec:sut} turned out to be representative,  and that we have 4 groups of VMs (4 first rows of Table~\ref{tab:vms_spec}), the bucket algorithm counts with $9 \times 4$ sets of buckets, one set for each transaction-group pair. Each set of buckets  comprises   $B=2$ buckets.


According to the cost function and the fault model considered in Section~\ref{ssec:modelbasedinsights}, the optimal bucket depth $D$ resides between 12 and 15. 
Table~\ref{tab:cs1_tbl} reports the performance metrics obtained from our experimental campaign, divided into two groups corresponding to $D=12$ and $D=15$. The first line of each group accounts for  {\it all} fault models, and the subsequent six lines correspond to the six fault models described in Section~\ref{ssec:toolsup}. 

\begin{table}[ht]
    \tiny
    \vspace{-0.1in}
    \centering
    \caption{Performance metrics, including Residual Effects counts (\textbf{RE}), \textbf{Pr}ecision, \textbf{Re}call and F-measure (\textbf{F1})  } 
    \vspace{-0.1in}
    \label{tab:cs1_tbl}
    \includegraphics[width=1\linewidth]{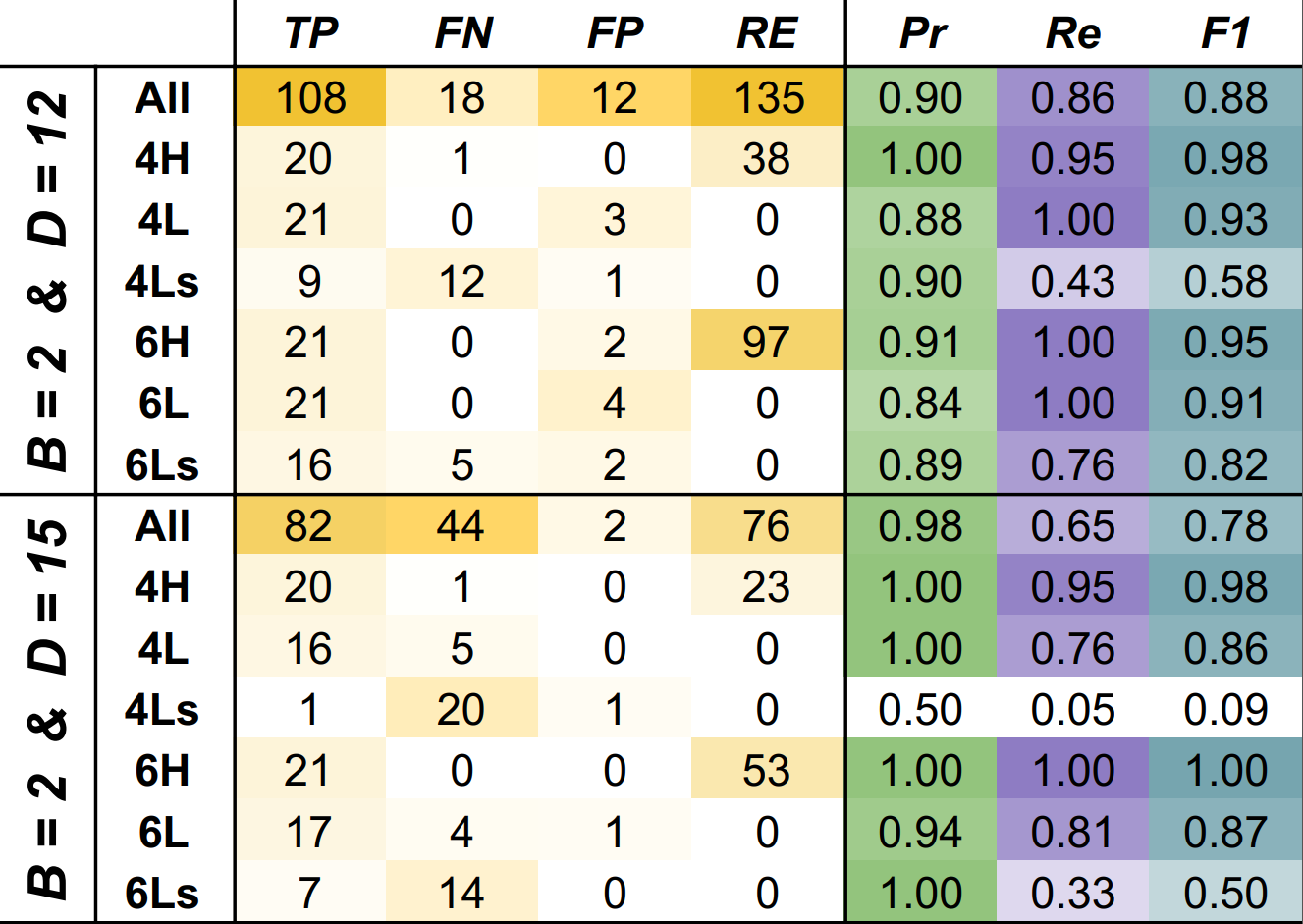}
\end{table}

Table~\ref{tab:cs1_tbl} indicates that the  proposed parametrization indeed yields a small number of false alarms  (column FP),  as suggested by the analytical model, and that the F-measure is typically greater than 0.78 (two notable exceptions being  under fault models \texttt{4Ls} and \texttt{6Ls}).  In addition,  Table~\ref{tab:cs1_tbl} also shows that  \gls{ba}  performance  varies as a function of the  anomaly   intensity and algorithm configuration. 
  In particular, letting $D=12$ or $D=15$ the algorithm is effective to detect high intensity anomalies (\texttt{4H} and \texttt{6H}) and low intensity anomalies of long duration (\texttt{4L} and \texttt{6L}).
  In those scenarios,  the  performance of the anomaly detector  under $D=12$ and $D=15$ is similar,   suggesting  robustness of the solution with respect to its parametrization.   
 
 To deal with   short anomalies of low-intensity (\texttt{4Ls} and \texttt{6Ls}), we  found that  $D$ must be fine tuned.   In the \texttt{6Ls} scenario, setting $D=12$  is key to control the number of  false negatives. Indeed,  $D=12$ produces  an  F-measure of 0.82 which significantly  outperforms an F-measure of  0.5 under $D=15$.    In the most challenging setup  \texttt{4Ls}, 
we must vary $D$ in a broader  range beyond 12 and 15  to detect  short bursts of faults.   Such observation, in turn, motivates a transaction-based parametrization  of the anomaly detector in those settings.  

In this work we consider the same value of $D$ for all transactions in order to show the effectiveness of the proposed approach in its simplest configuration.  Our preliminary results (not shown in this paper) indicate that  allowing a  distinct  parametrization  per-transaction   suffices to  deal with scenarios such as \texttt{4Ls}.  The decision between tuning the parameters in a  system wide manner or  in a per-transaction  basis must  balance between simplicity and  effectiveness, and we leave its  detailed experimental analysis as subject for future work.


\begin{figure}[htb]
\centering
\includegraphics[width=\linewidth]{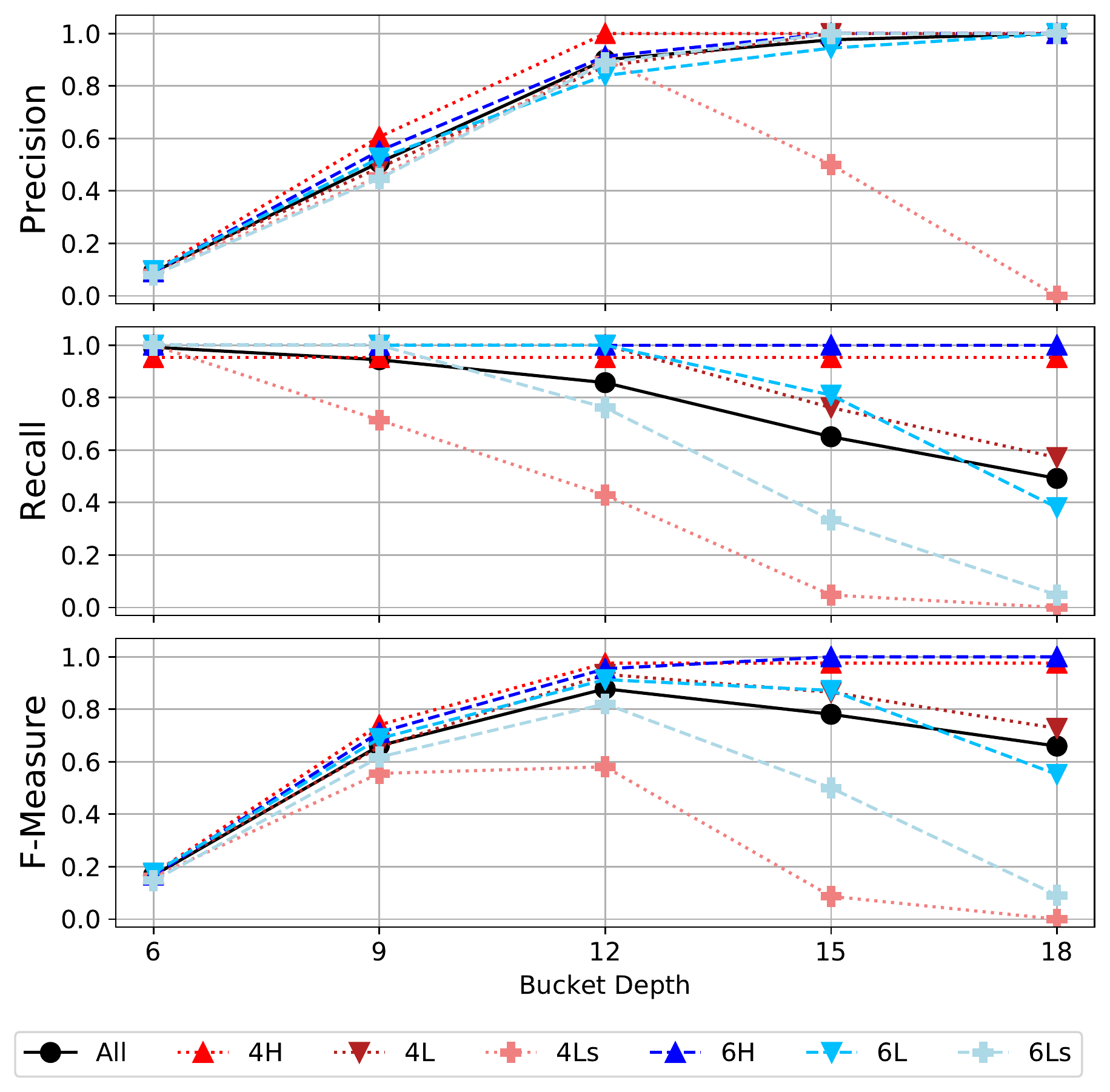}
\caption{Performance metrics  under the six  fault models. } 
\label{fig:all_metrics}
\end{figure}

 \figurename~\ref{fig:all_metrics} shows  how precision, recall and F-measure vary  as a function of $D$, for $D$ varying between 6 and 18.  
 Larger values of $D$ favor higher tolerance to performance variability under normal conditions.
  Therefore, the number of  \textit{false positives} (FPs) decreases and precision increases as  $D$ grows.  Recall (Re), in contrast,  decreases and the number of  \textit{false negatives} (FNs) increases as $D$ grows, as such growth produces  longer  detection  times. The F-measure  balances precision and recall, typically reaching its maximum value between $D=12$ and $D=15$ under the considered setups, which  is in agreement with the results obtained through the analytical  model parametrization in Section~\ref{ssec:modelbasedinsights}.  
  
\subsection{Take away message and engineering implications}
Whereas the model-based parametrization from Section~\ref{ssec:modelbasedinsights} is based on the cost function~\eqref{eq:cost1} to be minimized (Figure~\ref{fig:modelpar2}(b)), the F-measure yields an utility function to be maximized (Figure~\ref{fig:all_metrics}). 
In essence, \emph{  both the cost and  utility functions capture the fundamental tradeoff between detection time and false alarm rates} and are complementary to each other. 
While the experimental approach serves  for validation purposes and to explain system behavior in retrospect, after experiments are executed, the model-based approach  is key to perform what-if counterfactual analysis and for predictive purposes.

  

%


\section{Conclusion} 
\label{conc}

In this work, we presented a methodology for anomaly detection based on performance degradation, e.g.,  caused by security attacks at complex virtualized systems.  The approach leverages an analytical model to find the optimal parametrization of  an anomaly detector in a principled way. 

Our experimental assessment indicates the method's effectiveness by  injecting resource exhaustion attacks in a  virtualized system. Results show that it is possible to detect anomalous behavior using the throughput of the business transactions with an average precision of 90\% and recall of 86\%.  
Our experimental results also bring awareness about residual effects of  high-intensity fault loads, which may persist  after  active attacks have been interrupted. 

We believe that the  analytical and experimental contributions presented in this work advance the state of the art providing  novel perspectives towards the classical and fundamental tradeoff between detection time and false alarm rates faced in the optimal design of  anomaly detection mechanisms. 

For future research, we intend to extend our experiments with fault models representing other types of attacks, and to cope with a transaction-oriented system parametrization.



\section*{Acknowledgment}
This work was funded by CEFET-MG/Brazil, eSulab Solutions, CAPES, CNPq and FAPERJ, and by the Portuguese Foundation for Science and Technology (FCT) through the Ph.D. grant \texttt{SFRH/BD/144839/2019}, and the project METRICS (agreement no \texttt{POCI-01-0145-FEDER-032504}), within the scope of the project CISUC - UID/CEC/00326/2020 and by European Social Fund, through the Regional Operational Program Centro 2020. 
It was also supported SPEC RG Security Benchmarking (Standard Performance Evaluation Corporation; http://www.spec.org, http:// research.spec.org).


\bibliographystyle{IEEEtran}
\bibliography{IEEEabrv,references}


\end{document}